\documentclass[prd,preprintnumbers,twocolumn,aps]{revtex4-1}
\usepackage{graphicx}% Include figure files
\usepackage{dcolumn}% Align table columns on decimal point
\usepackage{bm}% bold math
\usepackage{epsfig,amsmath}
\usepackage{amssymb}
\usepackage{subfigure}
\usepackage[usenames,dvipsnames]{color}
\usepackage[pagebackref=false, colorlinks=true]{hyperref}
\definecolor{redish}{rgb}{0.7,0.2,0.0}  % color defined in (r=red,g=green,b=blue) model
\definecolor{bluish}{rgb}{0.2,0.5,0.8}

\hypersetup{linkcolor=redish,          % color of internal links
                  citecolor=blue,        % color of links to bibliography
                  filecolor=magenta,      % color of file links
                  urlcolor=bluish}          % color of the url

\DeclareFontFamily{U}{rsfs}{}         % Formal Script            %
\DeclareFontShape{U}{rsfs}{m}{n}{<5> rsfs5 <6><7> rsfs7          %
  <8><9><10><10.95><12><14.4><17.28><20.74><24.88> rsfs10}{}     %
\DeclareMathAlphabet{\mathfs}{U}{rsfs}{m}{n}

\def \O{\Omega}

\def \f{\frac}

\def \O{\Omega}

\def \e{\epsilon}
\def \p{\partial}

\def \th{\theta}

\begin{document}

\title{Spin precession in a black hole and naked singularity spacetimes}

\author{Chandrachur Chakraborty}
\email{chandrachur.chakraborty@tifr.res.in} 
\author{Prashant Kocherlakota}
\email{k.prashant@tifr.res.in}
\author{Pankaj S. Joshi}
\email{psj@tifr.res.in}
\affiliation{Tata Institute of Fundamental Research, Mumbai 400005, India}

\begin{abstract}
We propose here a specific criterion to address the existence
or otherwise of Kerr naked singularities, in terms of the precession 
of the spin of a test gyroscope due to the frame dragging by 
the central spinning body. 
We show that there is indeed an important characteristic difference in the 
behavior of gyro spin precession frequency in the limit of approach to these 
compact objects, and this can be used, in principle, to differentiate 
the naked singularity from black hole. 
Specifically, if gyroscopes are fixed all along the polar axis upto the horizon of a 
Kerr black hole, the precession frequency becomes arbitrarily high, 
blowing up as the event horizon is approached. On the other hand, in 
the case of naked singularity, this frequency remains always finite and 
well-behaved. Interestingly, this behavior is intimately related to and 
is governed by the geometry of the ergoregion in each of these cases which 
we analyze here. One intriguing behavior that emerges is, in the Kerr 
naked singularity case, the Lense-Thirring
precession frequency ($\O_{\rm LT}$) of the gyroscope due to 
frame-dragging effect decreases as
($\O_{\rm LT}\propto r$) after reaching a maximum, in the limit of $r= 0$,
as opposed to $r^{-3}$ dependence in all other known astrophysical cases.
\end{abstract}

\maketitle

\section{Introduction}
A rotating astrophysical compact object such as a Kerr black hole is characterized
by two parameters, namely the mass $(M)$ and angular momentum $(J)$. 
For the stationary vacuum Kerr solution when the dimensionless Kerr parameter
satisfies $a_{*}=J/M^2\leq 1$), then the Kerr singularity is enclosed within an
event horizon, which is a Kerr black hole. On the other hand, when $a_{*}>1$, the event horizon does not 
form and we get a Kerr naked singularity (also sometimes called a \lq superspinar\rq), 
which is visible in principle to external observers.

Such a formation or otherwise of a Kerr naked singularity in gravitational
collapse is an open issue. Also in other contexts the existence of Kerr naked
singularities has been a topic of intense discussion in recent years. 
For example, Jacobson and Sotiriou \cite{js} 
showed that a rotating black hole could be spun up from just below 
the extremal limit by capture of nonspinning test bodies if radiative and 
gravitational self-force (GSF) effects are neglected. Contrary to this, 
Barausse et al. \cite{bck} showed that it is possible to prevent this 
formation of naked singularities 
if one considers the conservative self-force. It was then shown by 
\cite{cb} that overspinning is possible (when the GSF is ignored)
only with particles thrown in from infinity. For any value of the initial energy 
at infinity, overspinning could be achieved by choosing the particle rest mass
and angular momentum from within certain open intervals. It was shown 
in Ref.\cite{w} that Kerr black holes are perturbatively stable and the Kerr
naked singularities
are unstable \cite{pani}. On the other hand, Saa and Santarelli \cite{ss} 
described a possible realistic scenario for the creation of a Kerr naked 
singularity from some recently discovered rapidly rotating 
black hole candidates in radio galaxies. Using the Polish doughnut model (not directly in the Kerr background), 
Li and Bambi \cite{lb} 
showed that the super-spinning compact objects can be generated by thick 
accretion disks. In order to investigate the observational testability of the Kerr 
bound, namely the possibility of $a_*>1$, Harada and Takahashi \cite{ht} 
calculated the energy spectra from an optically thick and geometrically thin 
accretion disk of a superspinning object and found that the observational 
confirmation of the Kerr bound is very hard, thus not ruling out the violation of this 
bound. They suggested that the violation of the bound might be detectable 
if the continuum X-ray spectrum of the disk is available.
Recently, Nakao et al. \cite{nakao} showed that an over-spinning body 
may produce a geometry close to that of a Kerr naked singularity around itself 
at least as a transient configuration.

These efforts and the current scenario imply that the existence or 
otherwise of a Kerr
naked singularity is a fundamental issue with important theoretical and 
observational consequences, since the black hole and naked singularity may 
have very different observational signatures. An attempt to find
such signatures was made by studying the accretion disks around the black
hole and naked singularity with the same masses
\cite{narayan1,narayan2}.
Interestingly, the thermal emission spectra of the accretion disks 
for each of these turn out to have quite different physical characteristics, which 
may help distinguish these objects from each other. 

The observable signatures of such a naked singularity may be 
checked in future if at all it exists, and to this purpose we need to identify 
the physical signatures distinguishing the black hole configuration from a 
naked singularity. 
In view of this current debate on the existence or otherwise of 
naked singularities, what we need is specific physical criteria so that if at all 
naked singularities formed in astrophysical scenarios, we could 
detect the same, or distinguish them from black holes.
To this purpose, we consider here the effect of frame-dragging \cite{lt} 
on the spin of a test gyroscope \cite{schiff1}, due to the rotation of stationary spacetime.
As is known, the Gravity Probe B satellite measured the precession rate
due to geodetic and frame-dragging effect relative to the Copernican system or
the `fixed star'  HR8703, also known as IM Pegasi, 
of a test gyro due to the rotation of the Earth \cite{ev}. 
In principle, the spin precession of a test gyro can be considered to be a 
useful experiment to test general relativity in our physical 
universe. 

Here we show that, using strong gravity gyroscopic precession,
one could in principle conclude whether a Kerr naked singularity can be 
ruled out as a realistic astrophysical object.
To this end, we show that the gyro precession frequency becomes 
arbitrarily high in the limit of approach to the ergoregion, thus diverging
on the ergosurface for a
Kerr black hole. On the other hand, the precession rate is finite and fully
regular even in the vicinity ($r\rightarrow 0$) of the Kerr naked 
singularity, when there is no ergoregion. In fact, the precession rate in this case 
goes as $\O_{\rm LT}\propto r$, and this is an adverse example of the frame-dragging
effect, because generally the dependence is $\O_{\rm LT}\propto r^{-3}$, for other 
compact objects. This scenario allows us to devise a thought experiment 
to rule out the existence of Kerr naked singularities if one never observes 
this $\O_{\rm LT}\propto r$ feature in the frame-dragging effect of a test gyro.

\section{Ergoregion and the Naked Singularity}
The event horizon and ergoregion give rise to several interesting phenomena 
in Kerr geometry. The nature of ergoregion is important to distinguish Kerr 
black hole from a naked 
singularity as we show, and we analyze the same here. We point out that
we can distinguish the Kerr naked singularities when they exist, from a black hole, 
in terms of the gyro spin precession frequency behavior.

The Kerr metric in Boyer-Lindquist coordinates \cite{bl} is,
\begin{eqnarray} \nonumber
ds^2=-\left(1-\f{2Mr}{\rho^2}\right)dt^2-\f{4Mar \sin^2\theta}{\rho^2}d\phi dt 
+\f{\rho^2}{\Delta}dr^2
\\ \nonumber 
+\rho^2 d\theta^2+\left( r^2+a^2+\f{2Mra^2 \sin^2\theta}{\rho^2}\right) \sin^2\theta d\phi^2 
\label{kerr} \nonumber
\\ 
\end{eqnarray}
Here $a$ is the Kerr parameter defined as $a=J/M$, which is the angular momentum 
per unit mass and,
\begin{equation}
 \rho^2=r^2+a^2 \cos^2\theta,        \\          \Delta=r^2-2Mr+a^2 .
\label{k2}
\end{equation}
In black hole case $a\leq M$, the outer and inner horizons are located 
at $r_{\pm}=M(1 \pm \sqrt{1-a_*^2})$ and boundaries of outer and inner ergoregions are, 
\begin{equation}
r^e_{\pm}=M (1\pm  \sqrt{1 -a_*^2 \cos^2\th}).
\end{equation}
Therefore, $r_+$ and $r_-$ coincide with $r_+^e$ and $r_-^e$ respectively at the 
pole. Away from pole to the equator the ergoregion increases 
and is maximum in size at the equator.

On the other hand, in the case of Kerr naked singularity ($a_*> 1$), 
horizons do not exist. 
A real value for radius of ergosurface can be obtained only 
in a certain range of $\theta$. From above equation, 
for all angles $\th$ with $1/a_* < \cos\th$, the ergoradii $r^e_{\pm}$ are 
imaginary and ergoregion does not exist for the spacetime region 
$0 \leq \th < \cos^{-1}\left(1/a_*\right)$. But for angular range
$\cos^{-1}\left(1/a_*\right)\leq\th\leq \pi/2$, it is present.
Thus the ergoregion shrinks towards equator as 
$a_{*}$ is larger. In principle, the outer ergoregion never vanishes completely,
even for $a_*\rightarrow \infty$, when it lies only in the equatorial plane, 
its limiting volume being zero. Interestingly, 
in these coordinates, the radius of the outer ergoregion along the equator remains the same
($r^e_+|_{\th=\pi/2}=2M$), whether it is a Kerr black hole or superspinar, 
$i.e.$ for entire range $0 < a_{*} < \infty$.

\begin{figure}
\subfigure[]{\includegraphics[width=2in,angle=0]{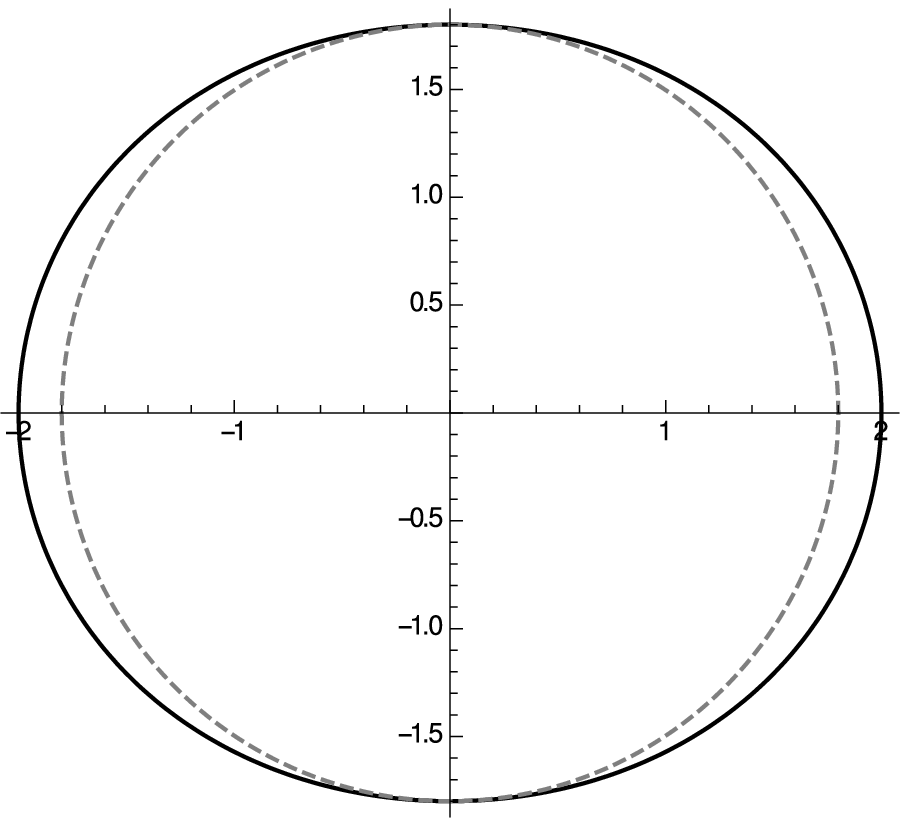}}
\hspace{0.05\textwidth}
\subfigure[]{\includegraphics[width=2in,angle=0]{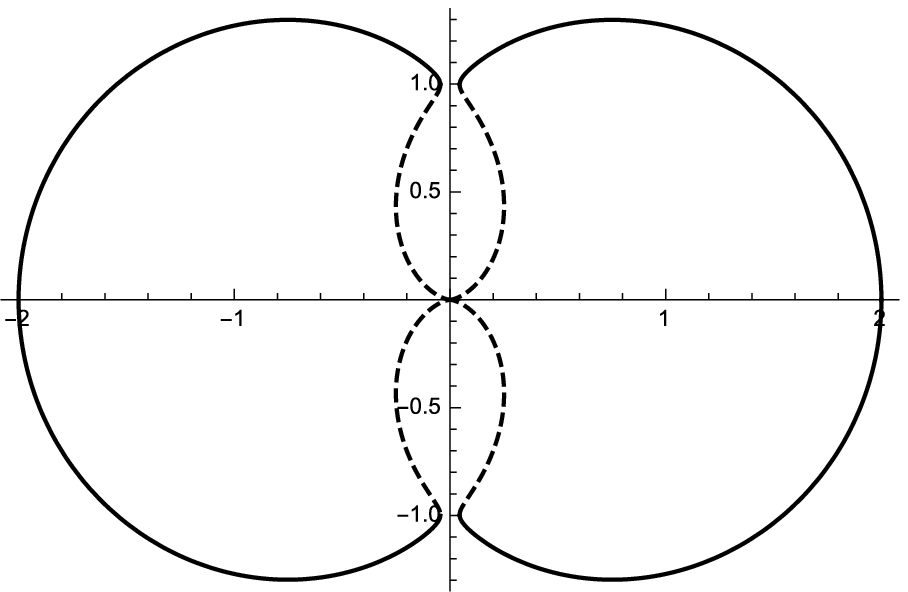}}
\hspace{0.05\textwidth}
\subfigure[]{\includegraphics[width=2in,angle=0]{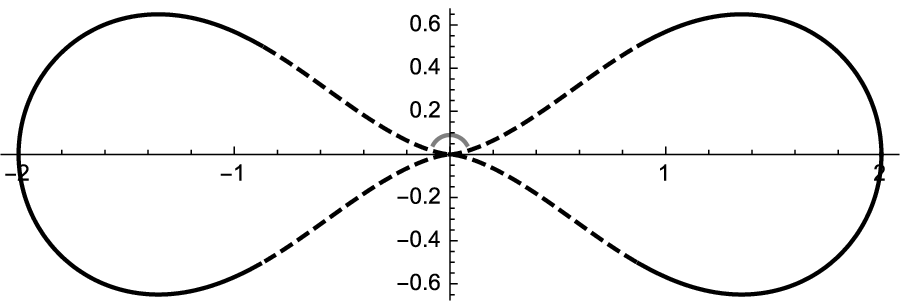}}
\caption{\label{fg2} Azimuthal sections of ergoregions for a Kerr black hole
and superspinar in Boyer-Lindquist coordinates. In $(a)$, the dashed line 
is outer event horizon and solid line is outer ergoradius ($a_{*}=0.6)$, ergoregion 
being in between. In $(b)$ and $(c)$, the solid/dashed lines
are outer/inner ergoradii for superspinar ($a_{*}=1.01$ and $a_{*}=2$). 
In $(c)$, for region of angle $2\pi/3$ there is no ergoregion.}
\label{ergo_BL}
\end{figure}

\begin{figure}
\begin{center}
\subfigure[]{
\includegraphics[width=2in,angle=0]{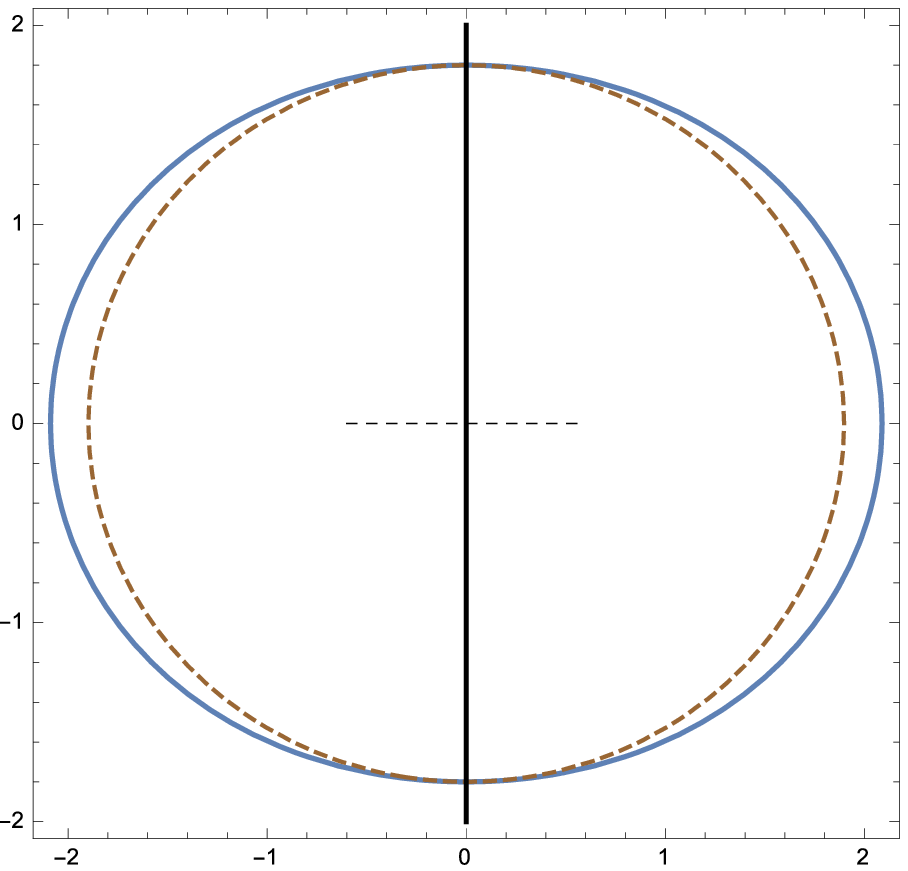}} 
\subfigure[]{
\includegraphics[width=2in,angle=0]{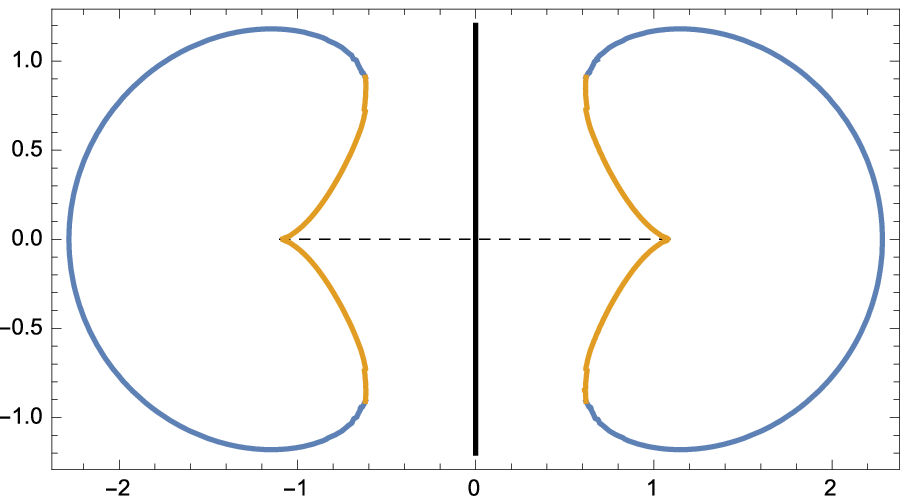}}
\subfigure[]{
\includegraphics[width=2in,angle=0]{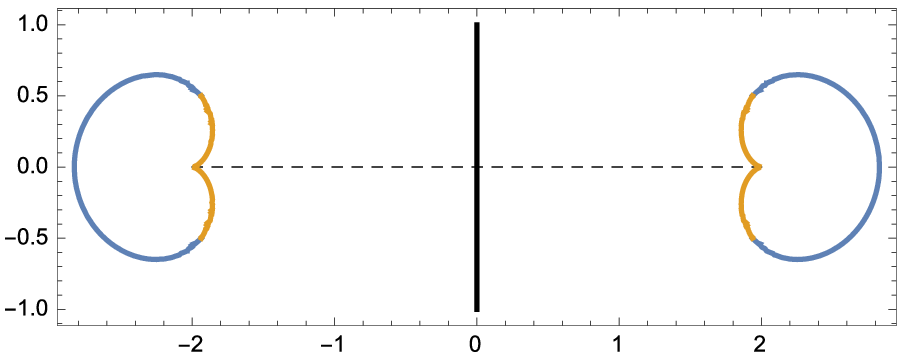}}
\caption{\label{fg3} Azimuthal sections of ergoregions for black hole and 
superspinar in Kerr-Schild
coordinates for same configurations as FIG.\ref{ergo_BL}. 
Same qualitative features better depicted, with orange/blue being inner/outer ergoradii,
brown is outer event horizon, and dashed line is ring singularity.}
\label{ergo_KS}
\end{center}
\end{figure}

We show the difference in ergoregion structure for a Kerr black 
hole versus superspinar 
in both Boyer-Lindquist and Cartesian Kerr-Schild coordinates \cite{kerr} 
in FIGs. \ref{ergo_BL} and \ref{ergo_KS}.
In  FIG.\ref{ergo_BL} (c), for $a_{*}=2$, the ergoregion exists for the
range $\pi/3\leq \th_e \leq \pi/2$ 
and $-\pi/3\geq \th_e \geq -\pi/2$, whereas there is no ergoregion for  
$-\pi/3 < \th_{ne} < \pi/3$. So the radii of the outer and inner ergoregions 
at the equator are $r^e_+=2M$ and $r^e_-=0$ respectively. 
The boundaries of inner and outer ergoregions coincide at $\th=\pi/3$ 
where the radii of both of the regions are $r^e_+=r^e_-=M$. 
In general, we can express the range of Boyer-Lindquist angles for which 
there is no ergoregion as,
\begin{eqnarray}
 -\cos^{-1}\f{1}{1+\e} < \th_{ne} < \cos^{-1}\f{1}{1+\e}
\end{eqnarray}
where $\e$ is the increment of $a_{*}$ from $1$ and it varies from 
$0<\e\leq \infty$ in principle. For a very small increment $\e$, for example 
$a_{*}=1.001$), the \lq opening angle\rq\ to access the vicinity of the ring 
singularity without crossing the ergoregion will be 
$\th_{op}(=2\th_{ne}) < 5.12^0$, because in this case we have   $-2.56^0<\th_{ne}<2.56^0$. 
Similarly, for the incremental values $\e=0.01,~0.1,~1$,
the values of the opening angle will be $\th_{op}<16.14^0,~49.24^0,~120^0$ respectively.

The above consideration shows that for any small increase in the
spin parameter beyond $a_*>1$, the 
structure of ergoregion changes in drastic manner. In particular, 
a finite angular cone opens up around the polar axis and we can place the gyros 
from outside to near $r=0$ through this region, without touching the ergoregion. 
On the other hand, since the ergoregion around a black hole covers it completely,
one must always pass through it when placing gyros from outside to near $r=0$. 
We show in the next section that the precession frequency of these test gyros 
diverges on the ergosurface and for a black hole,
this divergence would occur far before one reaches the vicinity of $r=0$. 
However, for the naked singularity spacetime 
one can place gyros all the way to $r=0$ in a broad angular region, 
with the gyro frequency remaining always finite and regular.

\section{Test gyros in Kerr spacetime}
The above structure of ergoregion helps us now to decide on existence 
of a naked singularity, or distinguish it from black hole, using the spin 
precession of a test gyroscope.  For our purpose, we can write 
the spin precession frequency of a test gyro due to the frame-dragging effect
relative to the Copernican system 
(or \lq fixed stars\rq), in a general stationary spacetime (see e.g.
\cite{str}),
\begin{equation}
\O_{\rm LT}= \frac{1}{2} \frac{\varepsilon_{ijl}}{\sqrt {-g}} \left[g_{0i,j}\left(\p_l - 
\frac{g_{0l}}{g_{00}}\p_0\right) -\frac{g_{0i}}{g_{00}} g_{00,j}\p_l\right]
\label{st}
\end{equation}
where $g_{o\mu}~(\mu=0,i)$ are the co-vector components of the timelike Killing
vector field $K=\p_0$ in a stationary spacetime, $g$ the determinant of metric
and $\varepsilon_{ijl}$ is Levi-Civita symbol. Eq.(\ref{st}) 
is derived assuming the gyro is held stationary and has four-velocity 
$u=(-K^2)^{-\frac{1}{2}}~K$. 
This means that gyros are held fixed everywhere in space and their precession
frequencies are analyzed to understand how they vary with change in spatial location.  
That is, the gyro remains along integral curves of $K$, which are not, in general, 
geodesics. Then the exact expression
for precession frequency of a test gyro due to frame-dragging effect
in Kerr spacetime was derived in \cite{cm} to
show that even in strong gravity regime this frequency is 
directly proportional to rotation of the spacetime and inversely proportional
to the cube of the distance. 
This precession frequency was calculated outside ergoregion in the black hole case,
and it diverges on the ergosphere. 
This precession frequency of a test gyroscope 
in a general Kerr spacetime, however, uses no weak gravity approximation
and is given from the above general formula as,
\begin{eqnarray}
\O_{\rm LT}(r,\theta)=\f{aM\left[4\Delta r^2 \cos^2\theta 
 +(\rho^2-2r^2)^2 \sin^2\theta \right]^\f{1}{2} }{\rho^3(\rho^2-2Mr)}. \nonumber
\label{krlt} 
\\
\end{eqnarray} 

\begin{figure}[h!]
\begin{center}
\subfigure[]{
\includegraphics[width=2.23in,angle=0]{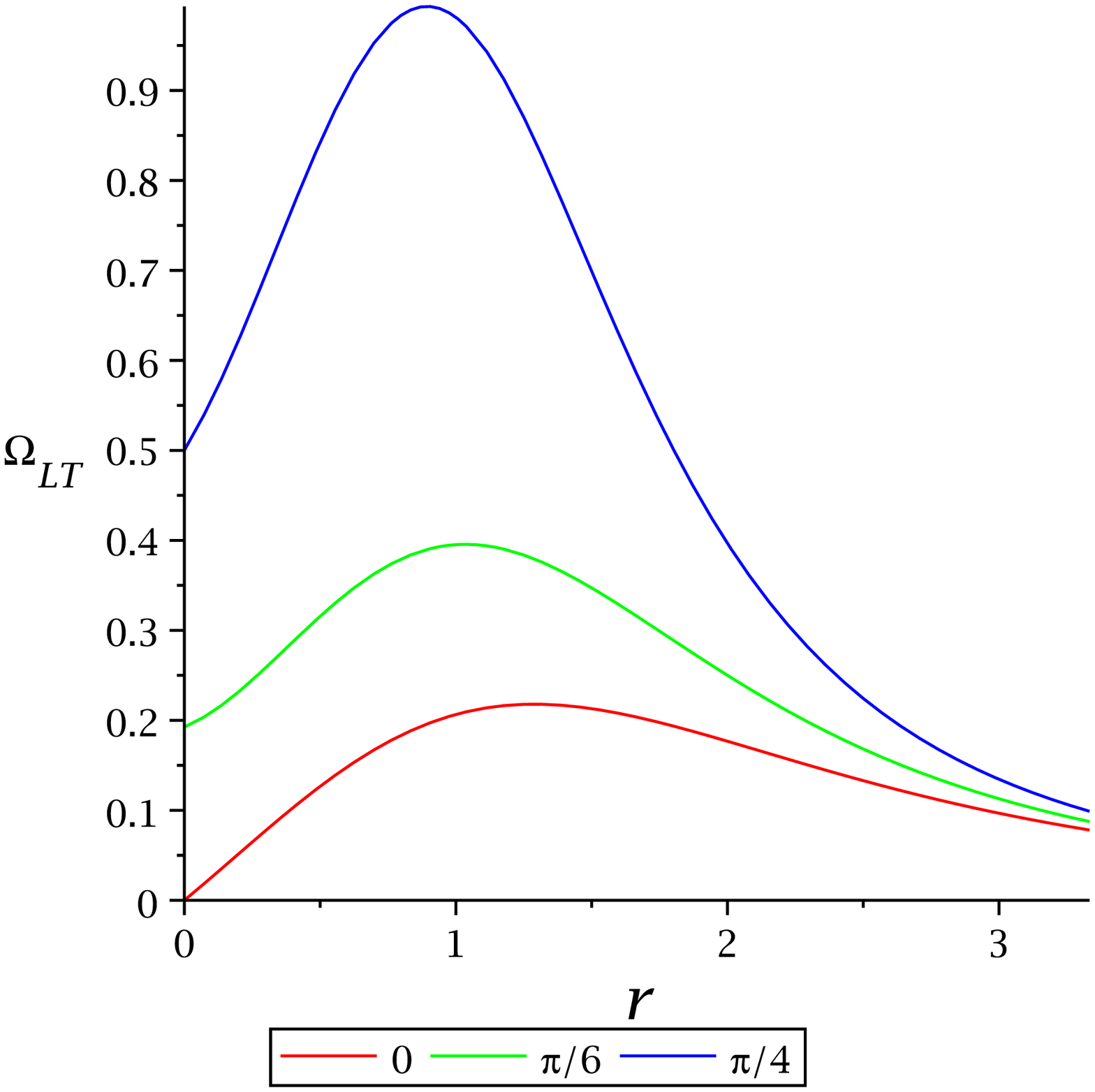}} 
\subfigure[]{
\includegraphics[width=2.23in,angle=0]{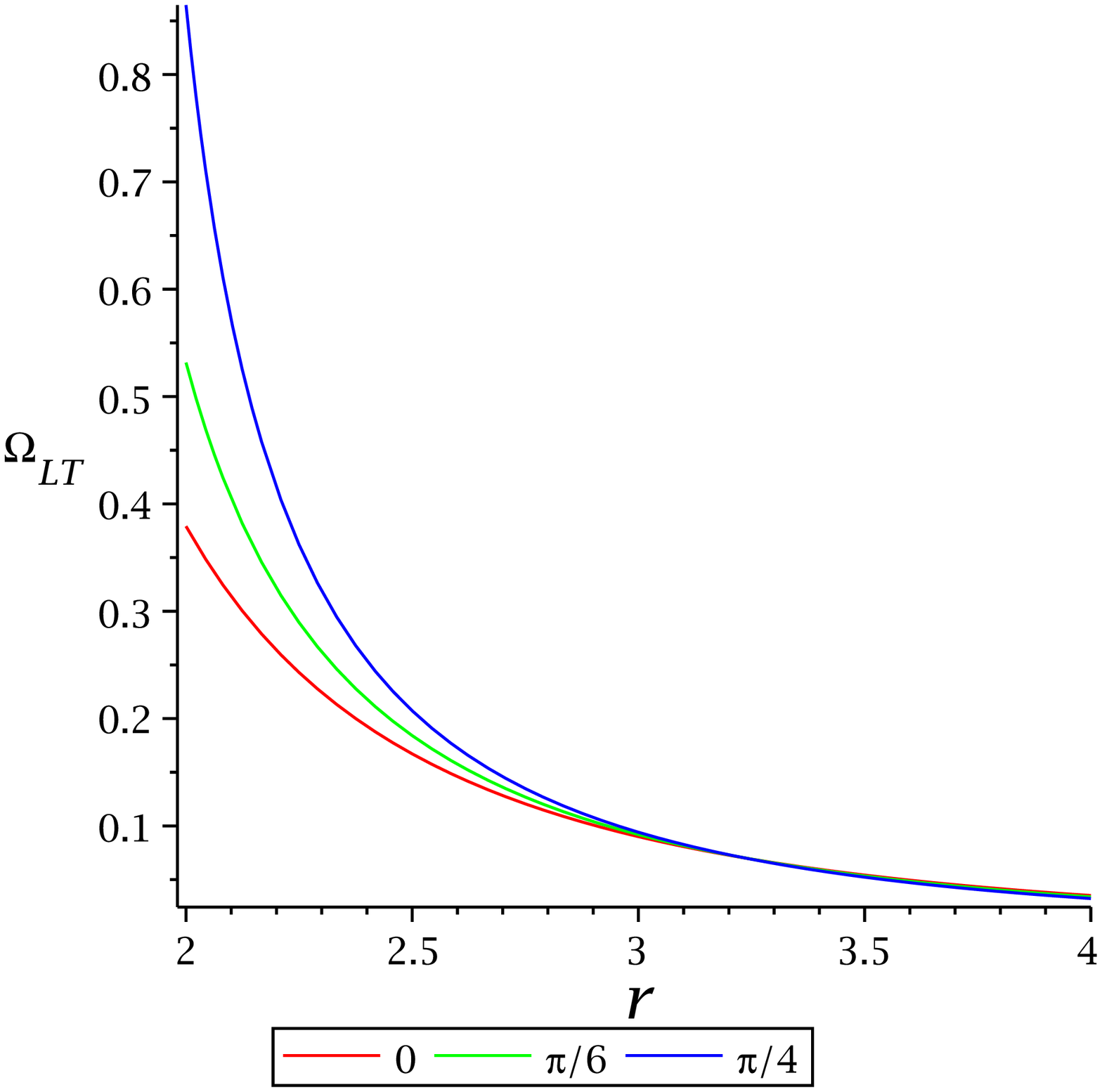}}
\subfigure[]{
 \includegraphics[width=2.23in,angle=0]{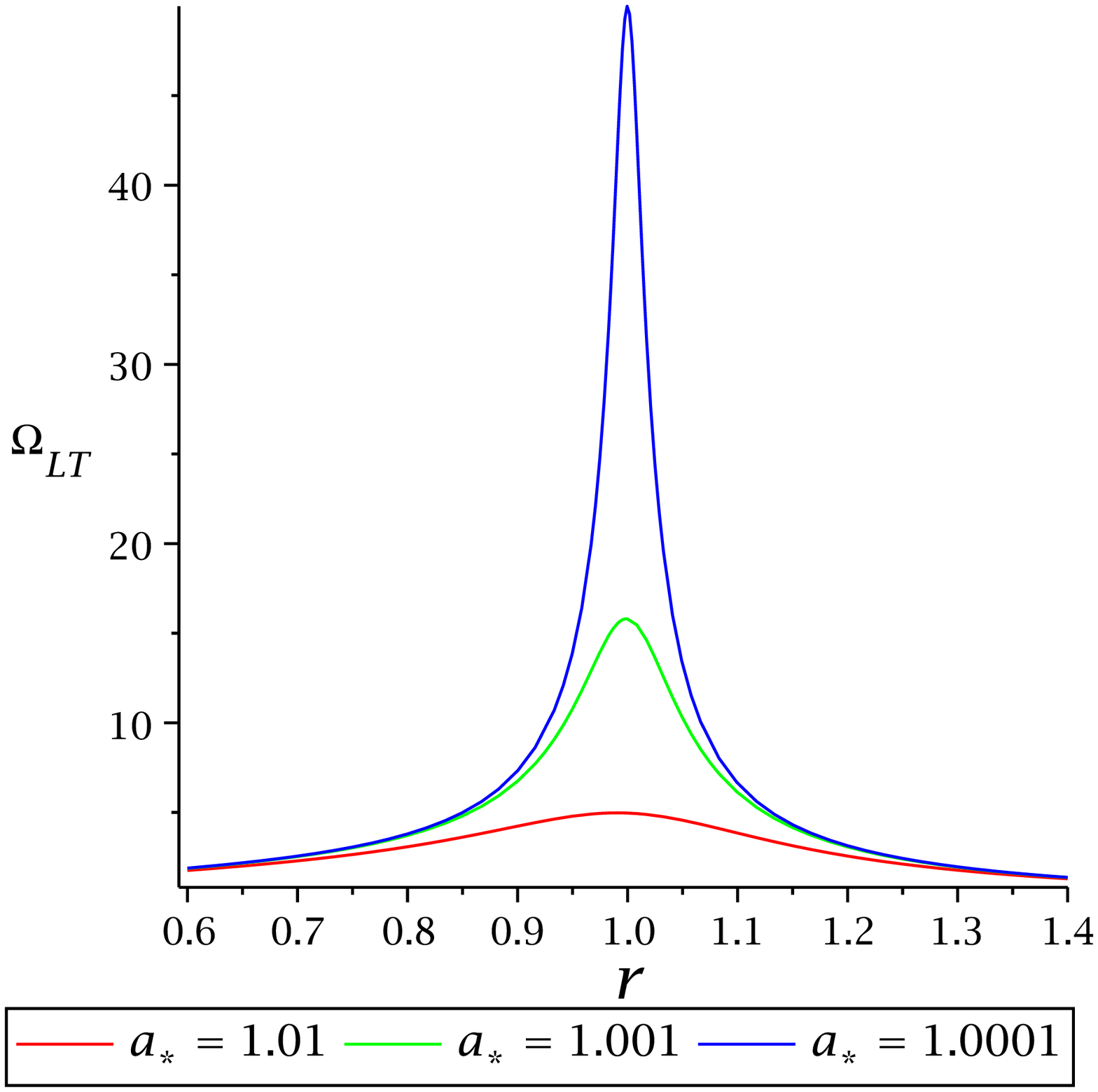}}
 \caption{\label{fg2} Variation  of $\O_{\rm LT}$ (in the unit of $M^{-1}$) versus 
 $r$ (in the unit of $M$). In $(a)$, $\Omega_{\rm LT}(r)$  
 plotted along different Boyer-Lindquist angles for  
 naked singularity ($a_*=2)$, and in $(b)$ for black hole with $a_*=0.9$. 
 In $(c)$, we plot $\Omega_{\rm LT}(r, \th=0)$ for superspinars with different
 spin parameters to highlight variation in peak values of $\O_{\rm LT}$.}
 \label{lt}
\end{center}
\end{figure}

Now, if we plot (see FIG.\ref{lt}) $\O_{\rm LT} (r,\th)$ (see Eq.(\ref{krlt})) versus 
$r$ for $a>M$ (Panel (a)),
we can see that the nature of frame-dragging effect for a superspinar is such that
the precession frequency does not blow up like the
black hole case (Panel (b)). It can also be seen from Eq.(\ref{krlt}), that $\O_{\rm LT}$
diverges
just near the ergosurface. Since the ergoregion completely surrounds Kerr black hole, 
spin precession frequency will diverge in all directions when the test gyro
approaches the same.
However, the frame-dragging frequency is finite and regular at all points 
outside the ergoregion, which does not cover the superspinar fully. So we find
that there are regions close to $r=0$ which are accessed 
without meeting the ergoregion, and where the precession frequency is
finite and regular. 
The frequency in the naked singularity case shows a \lq peak,\rq\, 
and its location $r_{p}$ 
is deduced by 
differentiating Eq.(\ref{krlt}), and setting $\f{d\O_{\rm LT}}{dr}|_{r=r_p}=0$. 

For the case of naked singularity, we then see for example that as the 
gyros are placed along the polar axis, coming in from infinity, where the 
ergoregion does not exist, the frequency increases to reach a maximum, 
and then decreases as $r\rightarrow 0$. 
This is new behavior in the sense that $\O_{\rm LT}$ does not follow the
inverse cube law of distance for known astrophysical objects in the region $r < a$,
and is a typical characteristic of Kerr naked singularities. It would be useful
to note here that if the gyros are placed anywhere in the region from 
the pole to the angle $\th_e$, the same finite behavior persists. 
However, the slope of the precession frequency curve increases 
and then diverges as we go to the angular range $\th_e \leq\th \leq \pi/2$, 
due to the presence of the ergoregion. 
The peak precession frequency increases with increasing angle and shifts toward 
$r=0$, as shown in Panel (a) of FIG.\ref{lt}. In contrast to this, the precession 
frequency diverges on the ergosurface for the same angles, as shown in 
Panel (b) of FIG.\ref{lt}.  In Panel (c) of FIG.\ref{lt}, we show how the peak precession 
frequency changes with $a$ along a particular angular direction 
(we consider here the direction along the pole) for superspinars. In the region
$\cos{\theta} \not\simeq 0$ and $r\ll a$, 
to leading order in $r/a$, from Eq.(\ref{krlt}) we get,
\begin{eqnarray}
\O_{\rm LT}(r,\th)|_{r \ll a}\approx \f{M}{a^3\cos^4\th}\left[4r^2
 +a^2 \sin^2\th \cos^2\th \right]^\f{1}{2}. \nonumber
\label{ra}
\\
\end{eqnarray}
For $\th=0$, this reduces to
\begin{eqnarray}
 \O_{\rm LT}\approx\f{2Mr}{a^3}.
 \label{pole}
\end{eqnarray}
Eq.(\ref{pole}) shows that  $\O_{\rm LT}\propto r$ and $\O_{\rm LT}\propto a^{-3}$ for $r\ll a$ for
the Kerr naked singularity spacetime, whereas the spin precession rate is $\O_{\rm LT}\propto a$ 
and $\O_{\rm LT}\propto r^{-3}$ for the Kerr black hole case. It is seen from Eq.(\ref{pole}) that 
$\O_{\rm LT}\rightarrow 0$ along the pole as $r\rightarrow 0$.
If $\th$ (see Eq.(\ref{ra})) increases from $\th=0$, the value of $\O_{\rm LT}$
also increases for ${r\rightarrow 0}$ and acquires a non-zero frame-dragging 
rate.

This is the main difference by which we can now distinguish 
between a Kerr naked singularity from a Kerr black hole. In the region 
$r\ll a$, the slope of $\O_{\rm LT}$ changes as $4Mr/a^3\cos^4\th \left[4r^2
 +a^2 \sin^2\th \cos^2\th \right]^{\f{1}{2}}$.
This slope is linear along the pole but increases with 
the angle. We note that the nature 
of the spin precession frequency due to frame-dragging effect 
for an extremal Kerr black hole is same as non-extremal case. 

Now, we can take test gyroscopes and place these at fixed spatial locations,
for a rotating compact object, which is either a Kerr black hole or a naked
singularity, along say its polar axis. Then the behavior of 
spin precession frequency due to the frame-dragging effect can be examined.
Of course, as we pointed out above, such an
experiment will work even if the approach to $r=0$ is along any other
directions within the cone around the polar axis, which is exterior to the 
ergoregion, even if placing the gyros along the pole could be simpler 
or preferable. Then there are two possibilities for us to observe:
(i) After crossing a certain region, the spin precession of the gyro will 
become arbitrarily high and diverge; or,
(ii) the test gyro will achieve a high precession rate, say a peak value, after crossing a 
certain distance, and then the precession rate will decrease as it further approaches 
the compact object. Finally, the precession frequency might even start vanishing.
One can then say that for Case (i) the spacetime 
is that of a Kerr black hole, but for Case (ii) a Kerr naked singularity 
is indicated instead. It follows that if the features signaling Case (ii) are 
never observed, 
then we can conclusively state that Kerr naked singularities do not exist 
or at least their abundance is extremely small.
\\

\section{Discussion}~
We have studied here the
features of spin precession of a gyroscope due to frame dragging effect
when placed in the vicinity of the black hole as well as naked singularity,
and we find a clear distinction, in principle, between
these compact objects. Specifically, 
to establish the spin precession frequency to be a possible distinguishing 
criterion, as a matter of principle we considered here gyroscopes attached
to the simplest observers possible, that is stationary observers. 
Since these observers are at fixed locations in space, the
velocities of these observers are given by $u=(-K^2)^{-\frac{1}{2}}~K$ 
(where $K=\p_0$) and we 
find that the spin precession frequency of gyroscopes for such observers,
caused purely due to frame dragging, always blows up on the boundary of
ergoregion. We note the differences in the structure of 
ergoregions around black holes and naked singularities and describe
how these structural differences allow us to show clear characteristic
features of the precession frequency to distinguish 
black hole from naked singularity. 

Our motivation here was to study the precession of spins of 
gyroscopes attached to stationary observers, that
is, to answer physical questions like `How will the gyroscope of an 
astronaut holding her spaceship at a constant distance from a Kerr 
compact object behave?', and `Would gyro measurements by her 
distinguish black holes and singularities?'. 
A caveat here would be of course, how realizable such paths or
configuration of gyros can be. It may be that from an observational perspective,
such paths may not be realizable in practice, even though we have clearly
shown, at least in principle, that there are key and essential theoretical 
differences that distinguish the black hole and naked singularity. 
It would be therefore useful to study this issue by varying the allowed
trajectories for the gyros.  

From such a perspective, in another work \cite{sp}, we have 
calculated the spin precession of a gyroscope which 
rotates in a circular orbit (that is, we ascribe a non-zero azimuthal
component to the four-velocity of the gyro), and we find that a similar
distinction can be made in terms of the spin precession frequency
between black hole and naked singularity spacetimes. However, in that case, 
the precession is not a consequence purely of frame dragging. For this
class of observers, a blow up of the precession frequency is obtained 
at the event horizon for black holes and no such feature is obtained 
for naked singularities. 
Therefore, we believe that spin precession studies around compact 
objects could shed light on their nature, namely whether or not 
they possess a horizon, and that it warrants a further analysis 
into understanding the reason behind these divergences.

We would like to mention that the consequences of such a frame dragging also 
manifest possibly at the observational levels. From such a perspective, 
we examined recently the effects of frame dragging or orbital plane 
precession in the properties of accretion disks around a black hole 
and naked singularity. We find that the orbital plane precession frequencies 
in an accretion disk show similar features to the spin precession frequency 
of gyroscopes that we reported here, due to the frame-dragging effect. 
That is, the orbital plane precession frequency increases continuously 
as one approaches the inner edge of a black hole accretion disk.
On the other hand, on moving inwards in an accretion 
disk around a Kerr naked singularity, this frequency attains a peak and 
starts decreasing as one approaches the inner edge of the disk, which
is the innermost stable circular orbit. For high enough spins, namely 
$a_* > 1.089$, beyond the Kerr bound $a_*=1$, this frequency even 
vanishes at $r_{0}=0.5625~a_*^2M$ before one actually reaches the 
inner edge of the accretion disk. Thus we observe a change in 
sign or the `sense' of orbital plane precession frequency. 
The details of this work have been reported recently in \cite{sp}.
\\

{\bf Acknowledgments:} We thank Mandar Patil and Ramesh Narayan 
for useful discussions on these topics. We also thank the referees for their 
valuable comments.

\end{document}